# Eshelbian dislocation mechanics: $\boldsymbol{J}$-, $M$-, and $\boldsymbol{L}$-integrals of straight dislocations[☆]

Markus Lazar[∗], Eleni Agiasofitou

*Department of Physics, Darmstadt University of Technology, Hochschulstr. 6, D-64289 Darmstadt, Germany*

**Abstract**

In this work, using the framework of (three-dimensional) Eshelbian dislocation mechanics, we derive the $\boldsymbol{J}$-, $M$-, and $\boldsymbol{L}$-integrals of a single (edge and screw) dislocation in isotropic elasticity as a limit of the $\boldsymbol{J}$-, $M$-, and $\boldsymbol{L}$-integrals between two straight dislocations as they have recently been derived by Agiasofitou and Lazar [Int. J. Eng. Sci. 114 (2017) 16–40]. Special attention is focused on the $M$-integral. The $M$-integral of a single dislocation in anisotropic elasticity is also derived. The obtained results reveal the physical interpretation of the $M$-integral (per unit length) of a single dislocation as the total energy of the dislocation which is the sum of the self-energy (per unit length) of the dislocation and the dislocation core energy (per unit length). The latter can be identified with the work produced by the Peach-Koehler force. It is shown that the dislocation core energy (per unit length) is twice the corresponding pre-logarithmic energy factor. This result is valid in isotropic as well as in anisotropic elasticity. The only difference lies on the pre-logarithmic energy factor which is more complex in anisotropic elasticity due to the anisotropic energy coefficient tensor which captures the anisotropy of the material.

## 1. Introduction

The present work lies on the realm of the Eshelbian dislocation mechanics which can be considered as the unification of incompatible elasticity theory of dislocations and the so-called configurational or Eshelbian mechanics. Configurational or Eshelbian mechanics, based on the pioneering works of Eshelby [11, 24], owns its establishment and great development to Gérard A. Maugin (1944–2016) whose valued and numerous contributions (see, e.g., [25, 26, 27]) have strongly influenced and advanced this field letting a remarkable and valuable scientific inheritance.

The $\boldsymbol{J}$-, $M$-, and $\boldsymbol{L}$-integrals are known concepts of configurational mechanics mainly in problems of cracks and cavities, and arise from the conservation or balance laws with respect to translation, scaling, and rotation transformations, respectively [1, 10, 14, 25, 26]. In dislocation theory, up to now, the $\boldsymbol{J}$-, $M$-, and $\boldsymbol{L}$-integrals of dislocations are not so popular and known as the corresponding ones in fracture mechanics and less is known about the physical meaning of these integrals. Only the fact that the $\boldsymbol{J}$-integral of dislocations is equivalent to the Peach-Koehler force is nowadays recognized and accepted (e.g., [3, 15, 29, 32]). As far as the $M$-integral of a single dislocation is concerned, Rice [29] was the first to study the $M$-integral when centered on the dislocation line in the framework of two-dimensional, compatible, linear, isotropic elasticity theory; and found that the $M$-integral is equivalent to the pre-logarithmic energy factor. This result, with a few physical merits, was used by Anderson, Hirth and Lothe [3], Asaro and Lubarda [4], and Baxevanakis and Giannakopoulos [8].

A dislocation (straight dislocation, dislocation loop) is an elementary line defect in a three-dimensional crystal and the carrier of plasticity in the crystal producing incompatibilities. Therefore, the appropriate framework for the description of dislocation problems is the three-dimensional, incompatible elasticity theory. Based on such a framework, Lazar and Kirchner [22] gave the general expressions of the $\boldsymbol{J}$-, $M$-, and $\boldsymbol{L}$-integrals of dislocations showing that these integrals are non-conserved ones, since source terms appear due to the dislocation density and plastic distortion tensors. Agiasofitou and Lazar [2] have recently derived the $\boldsymbol{J}$-, $M$-, and $\boldsymbol{L}$-integrals between two parallel straight dislocations revealing in this way their physical meaning and significance. In particular, they applied the general expressions of the $\boldsymbol{J}$-, $M$-, and $\boldsymbol{L}$-integrals of dislocations [22] to the specific problem of the interaction of two-parallel (edge and screw) dislocations and found that the $M$-integral between two straight dislocations (per unit length) is the half of the corresponding interaction energy between the two dislocations (per unit length) plus twice the corresponding pre-logarithmic energy factor, giving in this way to the $M$-integral, the physical interpretation of the interaction energy between the two straight dislocations. Moreover, they showed that the $L_3$-integral of two straight dislocations is the $z$-component of the configurational vector moment or the rotational moment (torque) about the $z$-axis caused by the interaction of the two dislocations.

In this work, we give a compact and systematic presentation of the $\boldsymbol{J}$-, $M$-, and $\boldsymbol{L}$-integrals of dislocations; starting from the general expressions of the $\boldsymbol{J}$-, $M$-, and $\boldsymbol{L}$-integrals (straight dislocations and dislocation loops) in Section 3, passing to the $\boldsymbol{J}$-, $M$-, and $\boldsymbol{L}$-integrals of

---

[☆]Dedicated to Gérard A. Maugin.
[∗]Corresponding author.
*Email addresses:* `lazar@fkp.tu-darmstadt.de` (Markus Lazar), `agiasofitou@mechanik.tu-darmstadt.de` (Eleni Agiasofitou)

straight dislocations in Section 4 and specifying to the interaction problem between two-parallel dislocations in isotropic elasticity in Section 5. In Section 6, the $J$-, $M$-, and $L$-integrals of a single dislocation (edge and screw) are derived as a limit of the $J$-, $M$-, and $L$-integrals between two straight dislocations in isotropic elasticity. The remarkable outcome is that the $M$-integral (per unit length) represents the total energy (per unit length) of the dislocation which is given by the sum of the self-energy (per unit length) and the dislocation core energy (per unit length). Section 7 is devoted to the derivation of the $M$-integral of a straight dislocation in anisotropic elasticity accompanied by important comparisons between the $M$-integral in anisotropic and isotropic elasticity. The expression of the dislocation core energy in anisotropic elasticity is explicitly given.

## 2. Fundamentals of linear incompatible elasticity theory of dislocations

The present work is based on the framework of three-dimensional, incompatible, linear elasticity theory of dislocations. Considering a homogeneous, linear, elastic body containing dislocations, the elastic energy density reads

$$W = \frac{1}{2} C_{ijkl} \beta_{ij} \beta_{kl} , \qquad (1)$$

where $\beta_{ij}$ is the non-symmetric *elastic distortion tensor* and $C_{ijkl}$ is the *tensor of the elastic constants* possessing the usual symmetries $C_{ijkl} = C_{klij} = C_{ijlk} = C_{jikl}$. The stress tensor is a symmetric tensor and is given by

$$\sigma_{ij} = \frac{\partial W}{\partial \beta_{ij}} = C_{ijkl} \beta_{kl} , \qquad (2)$$

which is the Hooke law for full anisotropy. Using Eq. (2), the elastic energy density (1) can be written in the form

$$W = \frac{1}{2} \sigma_{ij} \beta_{ij} . \qquad (3)$$

For dislocations, causing self-stresses, the equilibrium condition reads

$$\sigma_{ij,j} = 0 . \qquad (4)$$

In the theory of incompatible elasticity, the *total distortion tensor* $\beta_{ij}^{\mathrm{T}}$, which is defined as the gradient of the displacement vector $u_i$, can be decomposed into elastic and plastic parts (e.g., [17, 18, 28])

$$\beta_{ij}^{\mathrm{T}} := u_{i,j} = \beta_{ij} + \beta_{ij}^{\mathrm{P}} , \qquad (5)$$

where $\beta_{ij}^{\mathrm{P}}$ is the *plastic distortion tensor* or *eigendistortion*. The *dislocation density tensor* $\alpha_{ij}$ is given by (see, e.g., [17, 18])

$$\alpha_{ij} = -\epsilon_{jkl} \beta_{il,k}^{\mathrm{P}} \qquad \text{or} \qquad \alpha_{ij} = \epsilon_{jkl} \beta_{il,k} , \qquad (6)$$

where $\epsilon_{jkl}$ denotes the Levi-Civita tensor. The dislocation density tensor satisfies the Bianchi identity: $\alpha_{ij,j} = 0$.

## 3. The general expressions of the $J$-, $M$-, and $L$-integrals for dislocations

### 3.1. $J$-integral and Peach-Koehler force

The $J$-integral of dislocations is given by (see [2, 22])

$$J_k := \int_V \partial_j P_{kj} \, \mathrm{d}V = \int_V \epsilon_{kjl} \sigma_{ij} \alpha_{il} \, \mathrm{d}V , \qquad (7)$$

where the *Eshelby stress tensor of dislocations* in incompatible elasticity theory reads

$$P_{kj} = W \delta_{jk} - \sigma_{ij} \beta_{ik} . \qquad (8)$$

The integrand of the volume integral on the right hand side in Eq. (7) is nothing but the *Peach-Koehler force density*

$$f_k^{\mathrm{PK}} = \epsilon_{kjl} \sigma_{ij} \alpha_{il} , \qquad (9)$$

which is the material or configurational force density of a dislocation $\alpha_{il}$ interacting with a stress field $\sigma_{ij}$. The stress field $\sigma_{ij}$ may be caused by another defect. The $J$-integral has the physical meaning of the *Peach-Koehler force*

$$J_k = \int_V \epsilon_{kjl} \sigma_{ij} \alpha_{il} \, \mathrm{d}V = \mathcal{F}_k^{\mathrm{PK}} . \qquad (10)$$

### 3.2. $M$-integral and configurational work

The $M$-integral of dislocations, valid for a $d$-dimensional problem ($d \geq 2$), is given by (see [2, 22] for technical details)

$$M := \int_V \partial_j Y_j \, \mathrm{d}V = \int_V \left\{ x_k f_k^{\mathrm{PK}} - \frac{d-2}{2} \beta_{ij}^{\mathrm{P}} \sigma_{ij} \right\} \mathrm{d}V , \qquad (11)$$

where the *scaling flux vector of dislocations* in incompatible elasticity theory reads

$$Y_j = x_k P_{kj} - \frac{d-2}{2} u_k \sigma_{kj} , \qquad (12)$$

and $\delta_{kk} = d$. In field theory, the pre-factor $-(d-2)/2$ is called the *scaling* or *canonical dimension* of the vector field $u_k$ (see also [22]).

The integrand of the volume integral on the right hand side of Eq. (11) consists of configurational or material work density terms. The first term is the configurational work produced by the Peach-Koehler force density and the second term is the dislocation energy caused by a plastic distortion $\beta_{ij}^{\mathrm{P}}$ of a dislocation in the stress field $\sigma_{ij}$. Therefore, the $M$-integral of dislocations can be written as

$$M = \int_V \left\{ x_k f_k^{\mathrm{PK}} - \frac{d-2}{2} \beta_{ij}^{\mathrm{P}} \sigma_{ij} \right\} \mathrm{d}V = \mathcal{W}^{\mathrm{PK}} + (d-2) U_{\mathrm{d}} , \qquad (13)$$

where the *configurational work produced by the Peach-Koehler force density* is denoted by

$$\mathcal{W}^{\mathrm{PK}} = \int_V x_k f_k^{\mathrm{PK}} \, \mathrm{d}V \qquad (14)$$

and $U_\text{d}$ is the *dislocation energy* defined by

$$U_\text{d} := -\frac{1}{2} \int_V \beta^\text{P}_{ij} \sigma_{ij} \, \text{d}V \,. \tag{15}$$

The *interaction energy* $U_\text{int}$ between a dislocation with plastic distortion $\beta^\text{P}_{ij}$ and another dislocation with stress field $\sigma_{ij}$ is defined by $U_\text{int} = 2U_\text{d}$ (see, e.g., [28]). In the case of a single dislocation, $U_\text{d}$ represents the *self-energy* $U_\text{s}$ of the dislocation (see, e.g., [16, 28]).

### 3.3. *L*-integral and configurational vector moments

The ***L**-integral* of dislocations reads (see [2, 22])

$$L_k := \int_V \partial_l M_{kl} \, \text{d}V = \int_V \epsilon_{kji} \Big\{ x_j f^\text{PK}_i + \beta^\text{P}_{jl} \sigma_{il} \\ + [\beta_{jl} \sigma_{il} + \beta_{lj} \sigma_{li}] \Big\} \, \text{d}V \,, \tag{16}$$

where the *total angular momentum tensor of dislocations* in incompatible elasticity theory reads

$$M_{kl} = \epsilon_{kji} \big[ x_j P_{il} + u_j \sigma_{il} \big] \,. \tag{17}$$

The volume integral on the right hand side in Eq. (16) consists of three distinct configurational or material (rotational) vector moments. The first term is the configurational vector moment produced by the Peach-Koehler force density, the second term is the configurational vector moment caused by a plastic distortion $\beta^\text{P}_{ij}$ in presence of the stress field $\sigma_{il}$ and the third term is a configurational vector moment due to the material anisotropy of the solid under consideration.

Therefore, the ***L**-integral* of dislocations for an anisotropic material reads

$$L_k = \int_V \epsilon_{kji} \Big\{ x_j f^\text{PK}_i + \beta^\text{P}_{jl} \sigma_{il} + [\beta_{jl} \sigma_{il} + \beta_{lj} \sigma_{li}] \Big\} \, \text{d}V \,, \tag{18}$$

and the ***L**-integral* of dislocations for an isotropic material reduces to

$$L_k = \int_V \epsilon_{kji} \Big\{ x_j f^\text{PK}_i + \beta^\text{P}_{jl} \sigma_{il} \Big\} \, \text{d}V \,. \tag{19}$$

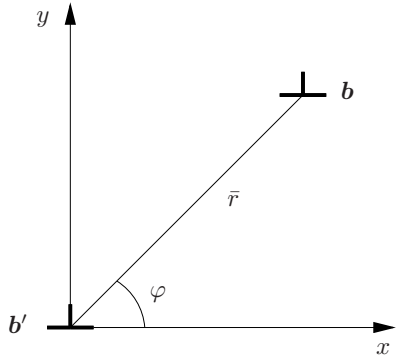

Figure 1: Interaction between two parallel straight dislocations with parallel Burgers vectors.

### 4. *J*-, *M*-, and *L*-integrals for straight dislocations

We give the ***J**-, *M*-, and *L*-integrals* of a straight dislocation with Burgers vector $\boldsymbol{b}$ at the position $(\bar{x}, \bar{y})$ (or in polar coordinates $(\bar{r}, \varphi)$) in the stress field of another straight dislocation with Burgers vector $\boldsymbol{b}'$ at the origin of the coordinate system $(0,0)$ (see Fig. 1) in a three-dimensional solid that means $d = 3$.

For a straight dislocation, the dislocation line $\mathcal{C}$ is a straight line and the dislocation surface $\mathcal{S}$ is a semi-infinite plane bounded by $\mathcal{C}$. We choose that $\mathcal{C}$ lies along the $z$-axis and $\mathcal{S}$ is the part of the $xz$-plane for positive $x$. On the surface $\mathcal{S}$, the displacement vector $\boldsymbol{u}$ possesses a jump $\boldsymbol{b}$ being the Burgers vector. Then, the only non-vanishing components of the plastic distortion and dislocation density tensors for a straight Volterra dislocation at the position $(\bar{x}, \bar{y})$ are given by

$$\begin{aligned}\beta^\text{P}_{iy}(x - \bar{x}, y - \bar{y}) &= -b_i \, H(x - \bar{x}) \, \delta(y - \bar{y}) \\ &= -b_i \, \delta(y - \bar{y}) \int_0^\infty \delta(x - \bar{x} - x') \, \text{d}x' \,,\end{aligned} \tag{20}$$

$$\alpha_{iz}(x - \bar{x}, y - \bar{y}) = b_i \, \delta(x - \bar{x}) \, \delta(y - \bar{y}) \,, \tag{21}$$

where $\delta(.)$ and $H(.)$ denote the Dirac delta function and the Heaviside step function, respectively. Note that the plastic distortion $\beta^\text{P}_{iy}$ possesses a discontinuity on $\mathcal{S}$.

If we substitute Eqs. (20) and (21) into Eqs. (10), (13), (18) and (19) and perform the volume integration, we obtain the ***J**-, *M*-, and *L*-integrals* of straight dislocations

$$J_k = \mathcal{F}^\text{PK}_k = \epsilon_{kjz} b_i \sigma_{ij}(\bar{x}, \bar{y}) \, l_z \,, \tag{22}$$

$$M = \mathcal{W}^\text{PK} + U_\text{d} = \bar{x}_k J_k + \frac{l_z}{2} \int_0^\infty b_i \sigma_{iy}(\bar{x} + x', \bar{y}) \, \text{d}x' \,, \tag{23}$$

$$L_k = \epsilon_{kji} \bar{x}_j J_i - l_z \int_0^\infty \epsilon_{kji} b_j \sigma_{iy}(\bar{x} + x', \bar{y}) \, \text{d}x' \\ + \int_V \epsilon_{kji} \big( \beta_{jl} \sigma_{il} + \beta_{lj} \sigma_{li} \big) \, \text{d}V \,, \tag{24}$$

$$L_k = \epsilon_{kji} \bar{x}_j J_i - l_z \int_0^\infty \epsilon_{kji} b_j \sigma_{iy}(\bar{x} + x', \bar{y}) \, \text{d}x' \,, \tag{25}$$

where $l_z$ is the length of the dislocation line. Eqs. (22)–(24) are valid for straight dislocations in anisotropic materials. Eq. (25) is valid for straight dislocations only in isotropic materials. In Eq. (22), we obtained the Peach-Koehler force formula for straight dislocations from the general expression of the *J*-integral. In Eqs. (23)–(25), it can be seen that the first terms appear due to the contribution of the Peach-Koehler force (or *J*-integral) and the second terms appear due to the plastic distortion of a straight dislocation. In Eq. (24), an additional term (volume integral) appears due to the anisotropy of the material. In Eqs. (23)–(25), the position vector $\bar{x}_k$ arises from the position of the dislocation density (21) due to the volume integration. For straight dislocations, the configurational work produced by the Peach-Koehler force reads

$$\mathcal{W}^\text{PK} = \bar{x}_k J_k = \bar{x}_k \mathcal{F}^\text{PK}_k \,. \tag{26}$$

# 5. $J$-, $M$-, and $L_3$-integrals of two parallel dislocations in isotropic elasticity

In this section, we give the $J$-, $M$-, and $L_3$-integrals of parallel edge dislocations with Burgers vectors in $x$-direction, parallel edge dislocations with Burgers vectors in $y$-direction, and parallel screw dislocations with Burgers vectors in $z$-direction. For detailed derivations readers are referred to Agiasofitou and Lazar [2].

## 5.1. Parallel edge dislocations with Burgers vectors in $x$-direction

For two parallel edge dislocations with Burgers vectors in $x$-direction, the $J$-, $M$-, and $L_3$-integrals per unit dislocation length are given by [2]

$$\frac{J_1}{l_z} = 2K^e_{xx} \frac{\cos\varphi \cos 2\varphi}{\bar{r}}, \tag{27}$$

$$\frac{J_2}{l_z} = 2K^e_{xx} \frac{\sin\varphi(2 + \cos 2\varphi)}{\bar{r}}, \tag{28}$$

$$\frac{M}{l_z} = K^e_{xx}\left[2 - \ln\frac{\bar{r}}{L} - \sin^2\varphi\right], \tag{29}$$

$$\frac{L_3}{l_z} = K^e_{xx} \sin 2\varphi, \tag{30}$$

where

$$K^e_{xx}(b_x, b'_x) = \frac{\mu b_x b'_x}{4\pi(1-\nu)} \tag{31}$$

is the *pre-logarithmic energy factor* for edge dislocations with Burgers vectors in $x$-direction.

Here, $\mu$ is the shear modulus, $\nu$ is the Poisson ratio, $L$ is the size of the dislocated body (or outer cut-off radius), $\bar{r} = \sqrt{\bar{x}^2 + \bar{y}^2}$ is the distance between the two dislocations, and $\varphi$ is the location angle of the dislocation with Burgers vector $b$ (see Fig. 1).

## 5.2. Parallel edge dislocations with Burgers vectors in $y$-direction

For two parallel edge dislocations with Burgers vectors in $y$-direction, the $J$-, $M$-, and $L_3$-integrals per unit dislocation length are given by [2]

$$\frac{J_1}{l_z} = 2K^e_{yy} \frac{\cos\varphi(2 - \cos 2\varphi)}{\bar{r}}, \tag{32}$$

$$\frac{J_2}{l_z} = -2K^e_{yy} \frac{\sin\varphi \cos 2\varphi}{\bar{r}}, \tag{33}$$

$$\frac{M}{l_z} = K^e_{yy}\left[2 - \ln\frac{\bar{r}}{L} + \sin^2\varphi\right], \tag{34}$$

$$\frac{L_3}{l_z} = -K^e_{yy} \sin 2\varphi, \tag{35}$$

where

$$K^e_{yy}(b_y, b'_y) = \frac{\mu b_y b'_y}{4\pi(1-\nu)} \tag{36}$$

is the *pre-logarithmic energy factor* for edge dislocations with Burgers vectors in $y$-direction.

## 5.3. Parallel screw dislocations

For two parallel screw dislocations with Burgers vectors in $z$-direction, the $J$-, $M$-, and $L_3$-integrals per unit dislocation length are given by [2]

$$\frac{J_1}{l_z} = 2K^s_{zz} \frac{\cos\varphi}{\bar{r}}, \tag{37}$$

$$\frac{J_2}{l_z} = 2K^s_{zz} \frac{\sin\varphi}{\bar{r}}, \tag{38}$$

$$\frac{M}{l_z} = K^s_{zz}\left[2 - \ln\frac{\bar{r}}{L}\right], \tag{39}$$

$$\frac{L_3}{l_z} = 0, \tag{40}$$

where

$$K^s_{zz}(b_z, b'_z) = \frac{\mu b_z b'_z}{4\pi} \tag{41}$$

is the *pre-logarithmic energy factor* for screw dislocations with Burgers vectors in $z$-direction.

## 5.4. Some highlights

The specific physical meaning of the $J$-, $M$-, and $L_3$-integrals of straight dislocations (Eqs. (27)–(40)) has been shown in [2]. In particular, the $J$-integral is the interaction force (Peach-Koehler force) between two dislocations; the $M$-integral between two straight dislocations per unit dislocation length is the half of the corresponding interaction energy per unit dislocation length plus twice the corresponding pre-logarithmic energy factor (configurational work done by the Peach-Koehler force being constant); and the $L_3$-integral of two straight dislocations is the $z$-component of the configurational vector moment or torque or rotational moment about the $z$-axis caused by the interaction of the two dislocations.

The $L_3$-integral for two parallel edge dislocations (see Eqs. (30) and (35)) is the torque depending on the angle $\varphi$ between the two dislocations. Therefore, the $L_3$-integral does have application for the interaction of two straight dislocations in isotropic elasticity in contrast to the claim of Anderson, Hirth and Lothe [3] that the $L_3$-integral has no application for dislocations in the "standard model".

For parallel edge dislocations, a $\sin^2\varphi$-part depending on the location angle $\varphi$ gives a contribution of angular dependence to the interaction energy in the $M$-integrals (29) and (34). Such a $\varphi$-dependence of the interaction energy between two parallel edge dislocations is of importance and recognized in the literature (see, e.g., [2, 5, 19]). In particular, this $\varphi$-dependence connects the $L_3$-integral with the $M$-integral [2]

$$L_3 = -\frac{\partial M}{\partial \varphi}. \tag{42}$$

Moreover, the $J$-integral is related with the $M$-integral as follows [2]

$$J_r = -2\frac{\partial M}{\partial \bar{r}}, \tag{43}$$

$$J_\varphi = -\frac{2}{\bar{r}}\frac{\partial M}{\partial \varphi}, \tag{44}$$

($J_r = J_1 \cos\varphi + J_2 \sin\varphi$ and $J_\varphi = J_2 \cos\varphi - J_1 \sin\varphi$). Combining Eqs. (42) and (44), a relation connecting the $J_\varphi$-, and $L_3$-integrals can be also obtained

$$J_\varphi = \frac{2}{\bar{r}} L_3. \tag{45}$$

## 6. $J$-, $M$-, and $L_3$-integrals of a single dislocation in isotropic elasticity

As it can be seen in Fig. 1, the limit from the interaction of two straight dislocations with Burgers vectors $\boldsymbol{b}$ and $\boldsymbol{b}'$ to the self-interaction of a singe dislocation with Burgers vector $\boldsymbol{b}$ can be obtained for $\bar{r} = \epsilon$ and $\varphi = 0$ in cylindrical coordinates and $\boldsymbol{b} = \boldsymbol{b}'$, which corresponds to $\bar{x} = \epsilon$ and $\bar{y} = 0$ in Cartesian coordinates. Here, $\epsilon$ is the inner cut-off radius being proportional to the constant dislocation core radius (see also [28]).

Therefore, the $J$-, $M$-, and $L_3$-integrals of a single dislocation can be obtained from Eqs. (27)–(41) for $\bar{r} = \epsilon$ and $\varphi = 0$ or $\bar{x} = \epsilon$ and $\bar{y} = 0$ and $\boldsymbol{b} = \boldsymbol{b}'$ in the corresponding expressions (see also Fig. 1).

### 6.1. Edge dislocation with Burgers vector in $x$-direction

For a single edge dislocation with Burgers vector $b_x$, we obtain from Eqs. (27)–(30), the $J$-, $M$-, and $L_3$-integrals per unit dislocation length, respectively

$$\frac{J_1}{l_z} = 2 K^e_{xx} \frac{1}{\epsilon}, \tag{46}$$

$$\frac{J_2}{l_z} = 0, \tag{47}$$

$$\frac{M}{l_z} = K^e_{xx} \left[ \ln\frac{L}{\epsilon} + 2 \right], \tag{48}$$

$$\frac{L_3}{l_z} = 0, \tag{49}$$

where

$$K^e_{xx}(b_x) = \frac{\mu b_x^2}{4\pi(1-\nu)} \tag{50}$$

is the *pre-logarithmic energy factor* for a single edge dislocation with Burgers vector in $x$-direction.

According to Eq. (23) (where $U_s$ has to be used for a single dislocation instead of $U_d$), Eq. (48) can be decomposed as follows

$$\frac{M}{l_z} = \frac{U_s}{l_z} + \frac{\mathcal{W}^{\text{PK}}}{l_z}, \tag{51}$$

where

$$\frac{U_s}{l_z} = K^e_{xx} \ln\frac{L}{\epsilon} \tag{52}$$

is the self-energy (per unit dislocation length) of an edge dislocation with Burgers vector in $x$-direction and

$$\frac{\mathcal{W}^{\text{PK}}}{l_z} = 2 K^e_{xx} \tag{53}$$

is the corresponding configurational work produced by the Peach-Koehler force being constant.

### 6.2. Edge dislocation with Burgers vector in $y$-direction

For a single edge dislocation with Burgers vector $b_y$, we obtain from Eqs. (32)–(35), the $J$-, $M$-, and $L_3$-integrals per unit dislocation length, respectively

$$\frac{J_1}{l_z} = 2 K^e_{yy} \frac{1}{\epsilon}, \tag{54}$$

$$\frac{J_2}{l_z} = 0, \tag{55}$$

$$\frac{M}{l_z} = K^e_{yy} \left[ \ln\frac{L}{\epsilon} + 2 \right], \tag{56}$$

$$\frac{L_3}{l_z} = 0, \tag{57}$$

where

$$K^e_{yy}(b_y) = \frac{\mu b_y^2}{4\pi(1-\nu)} \tag{58}$$

is the *pre-logarithmic energy factor* for a single edge dislocation with Burgers vector in $y$-direction.

### 6.3. Screw dislocation with Burgers vector in $z$-direction

For a single screw dislocation with Burgers vector $b_z$, we obtain from Eqs. (37)–(40), the $J$-, $M$-, and $L_3$-integrals per unit dislocation length, respectively

$$\frac{J_1}{l_z} = 2 K^s_{zz} \frac{1}{\epsilon}, \tag{59}$$

$$\frac{J_2}{l_z} = 0, \tag{60}$$

$$\frac{M}{l_z} = K^s_{zz} \left[ \ln\frac{L}{\epsilon} + 2 \right], \tag{61}$$

$$\frac{L_3}{l_z} = 0, \tag{62}$$

where

$$K^s_{zz}(b_z) = \frac{\mu b_z^2}{4\pi} \tag{63}$$

is the *pre-logarithmic energy factor* for a single screw dislocation with Burgers vector in $z$-direction.

**Remark:** The decomposition of the $M$-integral (51) is valid for edge and screw dislocations where in Eqs. (52) and (53) the corresponding pre-logarithmic energy factor has to be used.

### 6.4. Important outcomes

From Eqs. (46)–(49), (54)–(57) and (59)–(62), we make the following observations. For a single (edge or screw) dislocation in isotropic elasticity, the $J_1$-integral is proportional to the inverse dislocation core radius $1/\epsilon$; the $J_2$-, and $L_3$-integrals are zero, being conserved integrals; and the $M$-integral (per unit dislocation length) is equivalent to the corresponding self-energy (per unit dislocation length) plus twice the corresponding pre-logarithmic energy factor.

Alternatively, we can write the $M$-integral of a single straight dislocation in isotropic elasticity in a compact form independently of the edge or screw type, which is suitable for necessary comparisons with the corresponding $M$-integral in anisotropic elasticity which will be derived in the next section. Using the non-vanishing components of the so-called *energy coefficient tensor* $B_{ij}$ of isotropic elasticity in a Cartesian coordinate system whose $z$-axis is coincident with the dislocation line [6]

$$B_{xx} = B_{yy} = \frac{\mu}{4\pi(1-\nu)}, \qquad B_{zz} = \frac{\mu}{4\pi}, \qquad (64)$$

the $M$-integral (Eqs. (48), (56) and (61)) can be written in the following compact form[1]

$$\frac{M}{l_z} = b_i B_{ij} b_j \left[\ln\frac{L}{\epsilon} + 2\right] = E^{\text{is}}\left[\ln\frac{L}{\epsilon} + 2\right], \qquad (65)$$

where

$$E^{\text{is}} = b_i B_{ij} b_j \qquad (66)$$

is the *isotropic pre-logarithmic energy factor* with $B_{ij}$ given by Eq. (64). The decomposition of the $M$-integral (51) holds with the self-energy (52) (per unit dislocation length) to be written as

$$\frac{U_{\text{s}}}{l_z} = b_i B_{ij} b_j \ln\frac{L}{\epsilon} = E^{\text{is}} \ln\frac{L}{\epsilon} \qquad (67)$$

and the configurational work done by the Peach-Koehler force (see Eq. (53)) (per unit dislocation length) to be given by

$$\frac{\mathcal{W}^{\text{PK}}}{l_z} = 2\, b_i B_{ij} b_j = 2 E^{\text{is}}. \qquad (68)$$

Now, it is important to observe that the configurational work $\mathcal{W}^{\text{PK}}$ per unit dislocation length, Eq. (68), done by the Peach-Koehler force is equivalent to the *dislocation core energy* of a straight dislocation (per unit length) in an infinite medium given by (see [9, 12])

$$\frac{U_{\text{core}}}{l_z} = 2\, b_i B_{ij} b_j = 2 E^{\text{is}} = \frac{\mathcal{W}^{\text{PK}}}{l_z}. \qquad (69)$$

Finally, we arrive to the remarkable result that the $M$-integral (per unit length) of a single dislocation represents the total energy $U_{\text{total}}$ of the dislocation (per unit length) which consists of the self-energy (per unit length) plus the dislocation core energy (per unit length)

$$\frac{M}{l_z} = \frac{U_{\text{s}}}{l_z} + \frac{U_{\text{core}}}{l_z} = \frac{U_{\text{total}}}{l_z}. \qquad (70)$$

The total energy per unit length in Eq. (70) is in agreement with the total energy per unit length given by Bollmann [9] and Hull and Bacon [13].

Moreover, it is interesting to notice that the following fundamental relations hold

$$J_1 = J_r = -2\frac{\partial M}{\partial \epsilon} = -2\frac{\partial U_{\text{s}}}{\partial \epsilon}. \qquad (71)$$

The fact that the $J_1$-integral of a single dislocation is non-zero is a result valid in elasticity theory giving a singular stress at the dislocation line. In generalized elasticity theories like strain-gradient elasticity and nonlocal elasticity theories, the stress is zero at the dislocation line and, consequently, both components $J_1$ and $J_2$ are zero at the dislocation line (see, e.g., [20, 21, 23]). Thus, in non-singular dislocation theory, the $\boldsymbol{J}$-integral or the Peach-Koehler force of a single dislocation is zero at the dislocation line. However, only compatible elasticity theory in absence of defects and body forces for homogeneous materials is scale-invariant and within this framework the $M$-integral is conserved ($M = 0$). In the framework of incompatible, three-dimensional elasticity theory in presence of dislocations, it is shown that the $M$-integral represents the total energy ($U_{\text{s}} + U_{\text{core}}$) in the case of a single dislocation. Whereas in generalized elasticity theories even in absence of defects and body forces for homogeneous materials, the $M$-integral is not anymore conserved due to the appearance of length scales which break the scaling symmetry (see, e.g., [1, 22]).

## 7. $M$-integral of a single dislocation in anisotropic elasticity

In this section, we derive the $M$-integral of a single dislocation in the framework of anisotropic elasticity of dislocations as it is given by Bacon *et al.* [6]. We consider a single dislocation with Burgers vector $\boldsymbol{b}$ in an infinite, homogeneous, anisotropic, linear elastic solid with the tensor of the elastic constants $C_{ijkl}$ relative to fixed crystal axes. We use a fixed triad of mutually orthogonal unit vectors $\boldsymbol{m}$, $\boldsymbol{n}$, and $\boldsymbol{t}$ in the crystal so that $\boldsymbol{m} \times \boldsymbol{n} = \boldsymbol{t}$. Next, consider rotation of the $(\boldsymbol{m}, \boldsymbol{n}, \boldsymbol{t})$ vector system about the $\boldsymbol{t}$-axis, that is the vector $\boldsymbol{t}$ remains fixed. The orientation is given by an angle $\omega$ with respect to a reference system $(\boldsymbol{m}_0, \boldsymbol{n}_0, \boldsymbol{t})$. The dislocation line is running along the $\boldsymbol{t}$-axis. Let the vector $\boldsymbol{x}$ be orthogonal to $\boldsymbol{t}$; $\boldsymbol{x} \cdot \boldsymbol{t} = 0$. In addition, $\boldsymbol{m}$ and $\boldsymbol{n}$ are chosen so that the vector $\boldsymbol{m}$ is along the jump (or discontinuity) of the displacement field and the plastic distortion tensor. Then, the plastic distortion tensor is given by

$$\begin{aligned}\beta^{\text{P}}_{ij} &= -b_i n_j\, H(\boldsymbol{m}\cdot\boldsymbol{x})\,\delta(\boldsymbol{n}\cdot\boldsymbol{x}) \\ &= -b_i n_j\, \delta(\boldsymbol{n}\cdot\boldsymbol{x})\int_0^\infty \delta(\boldsymbol{m}\cdot\boldsymbol{x} - x')\,\text{d}x'.\end{aligned} \qquad (72)$$

We firstly calculate the self-energy $U_{\text{s}}$. Substituting Eq. (72) into Eq. (15) and performing the volume integration, we get

$$\frac{U_{\text{s}}}{l_z} = \frac{1}{2}\int_0^\infty \sigma_{ij} b_i n_j\, \text{d}x'. \qquad (73)$$

In Eq. (73) for $\boldsymbol{m}\cdot\boldsymbol{x} = |x| = |x'|$ and $\boldsymbol{n}\cdot\boldsymbol{x} = 0$, the stress tensor reads (see also [6, 7])

$$\sigma_{ij} = \frac{1}{2\pi|x|}\, C_{ijlp} b_s \big[ -m_p S_{ls} \\ + n_p (nn)^{-1}_{lk}\big[4\pi B_{ks} + (nm)_{kr} S_{rs}\big]\big], \qquad (74)$$

with the (anisotropic) *energy coefficient tensor* [6, 7]

$$B_{ij} = \frac{1}{8\pi^2}\int_0^{2\pi}\big[(mm)_{ij} - (mn)_{ir}(nn)^{-1}_{rk}(nm)_{kj}\big]\text{d}\omega \qquad (75)$$

---
[1] No summation in Eq. (65).

and the tensor

$$S_{ij} = -\frac{1}{2\pi} \int_0^{2\pi} (nn)_{ik}^{-1}(nm)_{kj}\, d\omega. \tag{76}$$

The tensors $B_{ij}$ and $S_{ij}$ depend on $\boldsymbol{t}$ and $C_{ijkl}$. Symbols of type $(ab)$ denote

$$(ab)_{ik} = a_j C_{ijkl} b_l. \tag{77}$$

The matrix $(nn)^{-1}$ is the inverse one of $(nn)$. Using the property [6]

$$\sigma_{ij} n_j = \frac{2}{|x|} B_{ij} b_j, \tag{78}$$

we obtain from Eq. (73) that

$$\frac{U_s}{l_z} = b_i B_{ij} b_j \int_0^\infty \frac{1}{|x'|}\, dx'. \tag{79}$$

In Eq. (79), if we replace the integral-limits 0 and $\infty$ by the inner cut-off radius $\epsilon$ and the outer cut-off radius $L$, respectively, we obtain the *self-energy of a single dislocation in anisotropic elasticity*

$$\frac{U_s}{l_z} = E^{\mathrm{an}} \int_\epsilon^L \frac{1}{|x'|}\, dx' = E^{\mathrm{an}} \ln \frac{L}{\epsilon} \tag{80}$$

with the *anisotropic pre-logarithmic energy factor* [6]

$$E^{\mathrm{an}} = b_i B_{ij} b_j \tag{81}$$

depending on $\boldsymbol{b}$, $\boldsymbol{t}$ and $C_{ijkl}$. Eq. (80) is in agreement with the dislocation self-energy in anisotropic elasticity given by Bacon *et al.* [6].

On the other hand, the configurational work (26) produced by the Peach-Koehler force can be written

$$\mathcal{W}^{\mathrm{PK}} = \bar{x}_1 \mathcal{F}_1^{\mathrm{PK}} + \bar{x}_2 \mathcal{F}_2^{\mathrm{PK}} = \bar{r} \mathcal{F}_r^{\mathrm{PK}}. \tag{82}$$

The component $\mathcal{F}_r^{\mathrm{PK}}$ of the Peach-Koehler force between two straight dislocations in anisotropic elasticity was given by Stroh [30] (see also [31]) and it reads in our notation as

$$\frac{\mathcal{F}_r^{\mathrm{PK}}}{l_z} = 2E^{\mathrm{an}} \frac{1}{\bar{r}}. \tag{83}$$

Next, taking the limit $\bar{r} = \epsilon$ in the Eqs. (82) and (83), we find that the *configurational work produced by the Peach-Koehler force of a single dislocation in anisotropic elasticity*

$$\frac{\mathcal{W}^{\mathrm{PK}}}{l_z} = \epsilon \Big(2E^{\mathrm{an}} \frac{1}{\epsilon}\Big) = 2E^{\mathrm{an}}, \tag{84}$$

*equals twice the anisotropic pre-logarithmic energy factor.*

It is noteworthy to see that the self-energy $U_s$ and the configurational work $\mathcal{W}^{\mathrm{PK}}$ (per unit length) of a single dislocation are given formally by the same expressions for isotropic (Eqs. (67) and (68)) and anisotropic (Eqs. (80) and (84)) elasticity. The difference lies on the pre-logarithmic energy factor which captures the anisotropy of the material through the energy coefficient tensor $B_{ij}$.

Finally, Eq. (23) in the case of a single dislocation using Eqs. (80) and (84) provides the *M-integral (per unit length) of a single dislocation in anisotropic elasticity*

$$\frac{M}{l_z} = \frac{U_s}{l_z} + \frac{\mathcal{W}^{\mathrm{PK}}}{l_z} = E^{\mathrm{an}}\Big(\ln \frac{L}{\epsilon} + 2\Big). \tag{85}$$

Eq. (85) states that the $M$-integral per unit length of a single dislocation in anisotropic elasticity is equivalent to the self-energy per unit length of the dislocation plus twice the anisotropic pre-logarithmic energy factor.

The result from isotropic elasticity that the configurational work $\mathcal{W}^{\mathrm{PK}}$ done by the Peach-Koehler force represents the dislocation core energy can be naturally taken over to anisotropic elasticity. Then, we can derive from the configurational work $\mathcal{W}^{\mathrm{PK}}$ (see Eq. (84)) the expression of the *dislocation core energy (per unit length) in anisotropic elasticity*,

$$\frac{U_{\mathrm{core}}}{l_z} = 2E^{\mathrm{an}} = 2\, b_i B_{ij} b_j, \tag{86}$$

which equals twice the anisotropic pre-logarithmic energy factor $E^{\mathrm{an}}$. Thus, the $M$-integral (85) gives the total energy of a dislocation in an anisotropic solid.

## 8. Conclusion

The main conclusions that can be reached are the following: Firstly, the physical interpretation of the $M$-integral of a single dislocation, which is equivalent to the total energy of the dislocation, which is given by the sum of the self-energy and the dislocation core energy. This result holds in isotropic as well as in anisotropic elasticity. Secondly, the dislocation core energy is twice the corresponding pre-logarithmic energy factor. This result is valid in isotropic (see Eq. (69)) as well as in anisotropic elasticity (see Eq. (86)); the only difference lies on the corresponding pre-logarithmic energy factor which is more complex in anisotropic elasticity due to the anisotropic energy coefficient tensor (75).

## Acknowledgments

The authors gratefully acknowledge a grant from the Deutsche Forschungsgemeinschaft (Grant No. La1974/4-1).